\documentclass[prd,twocolumn,showpacs,preprintnumbers,amsmath,amssymb,nofootinbib]{revtex4}
\usepackage{graphicx}
\usepackage{psfrag}

\DeclareMathOperator{\Tr}{Tr}
\DeclareMathOperator{\Real}{Re}
\DeclareMathOperator{\Imag}{Im}
\DeclareMathOperator{\vol}{vol}
\DeclareMathOperator{\arccosh}{arccosh}
\newcommand{\norder}[1]{:\!#1\!:}
\newcommand{\maxp}{cutoff}
\newcommand{\eqn}[1]{(\ref{#1})}

\begin{document}
\title{AdS/CFT description of D-particle decay}
\preprint{AEI-2004-038, hep-th/0405125}
\author{Kasper Peeters and Marija Zamaklar}
\affiliation{MPI f\"ur Gravitationsphysik,
  Am M\"uhlenberg 1,
  14476 Golm, GERMANY}
\email[Email:]{kasper.peeters,marija.zamaklar@aei.mpg.de}
\date{May 13th, 2004}
\begin{abstract}
Unstable D-particles in type-IIB string theory correspond to sphaleron
solutions in the dual gauge theory. We construct an explicit
time-dependent solution for the sphaleron decay on $S^3 \times
\mathbb{R}$, as well as the coherent state corresponding to the decay
product. We develop a method to count the number of bulk particles in
the AdS/CFT setup.  When applied to our coherent state, the naive
number operator $\hat{O}_J^\dagger\hat{O}^{\phantom{\dagger}}_J$ is
shown to be inappropriate, even in the large-$N$ limit. The reason is
that the final state consists of a large number of particles.  By
computing all probabilities for finding multi-particle states in the
coherent state, we deduce the bulk particle content of the final state
of the sphaleron decay.  The qualitative features of this spectrum are
compared with the results expected from the gravity side, and
agreement is found.
\end{abstract}
\maketitle

\section{Introduction and summary}
The spectrum of type-IIB string theory contains, apart from the stable
BPS and non-BPS states, also a wide variety of unstable
D-branes. These unstable branes contain a tachyon field on their
world-volume, and the condensation of this field corresponds to the
decay of the brane. Recently, a lot of progress has been made in
understanding the dynamical aspects of the decay of unstable branes.
Most of the analysis was performed directly using boundary conformal
field theory in flat space, initiated by Sen's construction of the
boundary states for decaying D-branes~\cite{Sen:2002nu}, or by using
the~$c=1$ matrix model for the description of the decay of D-branes in
1+1 dimensional string theory~\cite{McGreevy:2003kb}. In the present
paper we would like to study the problem of decaying branes in the
set-up of the ``standard'' AdS/CFT correspondence.

As was argued by Harvey et al.~\cite{Harvey:2000qu}, the unstable
D-branes in string-theory are equivalents of ``sphalerons'': they are
unstable solutions located at a saddle point of the potential in
configuration
% V2: added reference to Manton.
space, at the top of a non-contractible loop~\cite{Manton:1983nd}.  In
the context of the AdS/CFT conjecture, this correspondence between
unstable D-branes and sphalerons is in fact even more direct. It has
been argued by Drukker et al.~\cite{Drukker:2000wx} that the existence
of sphaleronic saddle points in the potential of the theories on both
sides is a feature which is preserved when going from weak to strong
coupling, despite the fact that the precise form of the potential
receives quantum corrections.  The unstable D-particles of string
theory are then in precise correspondence with known sphaleron
solutions of the dual gauge theory. Kinematical aspects of this
correspondence were investigated in detail in~\cite{Drukker:2000wx}.

In the present paper, we will analyse dynamical aspects of this
correspondence. Our analysis consists of three parts. First, we will
construct the classical solution of the decaying sphaleron, and obtain
a quantum mechanical description of the final stage of this decay
using a coherent state. We then develop the formalism to count the
number of bulk particles into which this final state decomposes. In
the last part, we apply this formalism to a concrete case at
finite~$N$ and extract qualitative features of the decay
process. Although the whole process is highly non-supersymmetric, and
thus expected to be subject to quantum corrections, we will see that
there are indeed qualitative features which agree with known results
derived on the string theory side.\footnote{This is to a certain extent
similar to the attempts to match the values of the entropy for AdS
black holes, using a calculation of the free energy in free
Yang-Mills theory \cite{Gubser:1998nz}. The main difference with
respect to this case, however, is in the dependence on the coupling
constant. The leading value of the black hole entropy and the free
Yang-Mills energy are independent of the string/YM coupling.
As we will see, the number of particles produced in the sphaleron
decay does depend on the coupling.} 
\bigskip

The dual gauge theory system is studied by considering a time
dependent solution of the decaying sphaleron on the three-sphere. We
are able to find an analytic, classical solution for the spherically
symmetric decay channel of the sphaleron.  While the non-abelian
character of the gauge theory (i.e.~the non-vanishing coupling) is
crucial for the existence of the sphaleron solution near the top of
the potential, it turns out that our solution abelianises near the
bottom of the potential valley (i.e.~it is a solution to the free
Yang-Mills equations of motion on the sphere).  This allows us to
construct a coherent state corresponding to the \emph{final} product
of the sphaleron decay.\footnote{Similar descriptions of the Standard
Model sphaleron decay using a coherent state approach have been
discussed by Zadrozny~\cite{Zadrozny:1992yf} and Hellmund and
Kripfganz~\cite{Hellmund:1992ub}.}  This coherent state should be dual
to the gas of closed string particles which is the decay product of an
unstable D-particle.

In order to make a link with calculations on the gravity side, we then
calculate the ``number operators'' in this coherent state $|c \rangle$
for various single trace operators~$\hat O_J$ which are dual to closed
string particles.  There are two subtle points in this procedure. One
is related to the fact that in gravity calculations one uses the
``standard'' notion of particles in the bulk as (angular) momentum
eigenstates, and calculates emission amplitudes for these particles.
Hence, in order to make a comparison with gravity possible, we cannot
directly use the AdS/CFT correspondence in position space. Instead, we
first have to construct boundary operators that are dual to bulk
angular momentum eigenstates. To construct these operators one
projects the composite operators onto eigen angular momentum
operators, by multiplying them with the appropriate tensorial
spherical harmonics and integrating over the sphere. This construction
is explained in the section~\ref{s:AdSparticles} and further
illustrated on an explicit example in appendix~\ref{a:s2opstate}.

The second subtlety in counting particles in the coherent state is
related to the fact that the operators $\hat{O}_J$ which create
elementary bulk particles are, from the point of view of the gauge
theory, \emph{composite} rather than elementary operators. The naive
number operator $\hat{O}^\dagger_J \hat{O}^{\phantom{\dagger}}_J$
turns out to be inappropriate; we will see that this is because it
only behaves as a counting operator when both $N\rightarrow\infty$ and
the number of particles $p$ in the state satisfies $p\ll N$.
Therefore, in order to count the number of particles corresponding to
an operator $\hat{O}_J$, one needs to calculate the probabilities
${\cal P}_p$ for finding a $p$-particle state individually. Since
$\hat{O}_J$ particles can appear in combination with any arbitrary
other (multi-particle) operator $\hat{O}_K$, the expression for
finding a $p$-particle $\hat{O}_J$ state is
% V2: states written differently to address issues concerning
%     orthonormalisation of the states.
\begin{equation}
\label{proba}
{\cal P}_p =\sum_{\hat{O}_K} 
  \frac{\displaystyle\left|\Big\langle (\hat{O}_J)^p \hat{O}_K\, \Big|\, c \Big\rangle \right|^2}{%
        \displaystyle
        \Big\langle (\hat{O}_J)^p \hat{O}_K\,\Big|\, 
                     (\hat{O}_J)^p \hat{O}_K  \Big\rangle 
        \langle c | c \rangle} \, .
\end{equation}
To see why computing~\eqn{proba} is hard, consider the simplest terms
in the sum, when $\hat{O}_K$ is just the identity operator. This term
is
% V2: see previous remark.
\begin{equation}
\label{e:introenergies}
\frac{\displaystyle \left|\Big\langle (\hat{O}_J)^p\, \Big|\, c \Big\rangle \right|^2}{%
     \displaystyle
     \Big\langle (\hat{O}_J)^p \,\Big|\, (\hat{O}_J)^p\, \Big\rangle  
  \langle c | c \rangle}
= \frac{\Big|(O_{J})^p\, \langle 0 | c \rangle\Big|^2 }{p! \left(1 + \frac{b(p,J)}{N^2} + \ldots\right)} \,  ,
\end{equation}
where~$O_J$ without a hat denotes the (positive frequency part of the)
classical expectation value of the operator~$\hat{O}_J$.  When~$b\ll
N$, the expression~\eqn{proba} can be summed, yielding the result one
would obtain using the naive number operator.  However, as indicated,
the coefficient~$b$ in the denominator depends on~$p$ and~$J$, and for
large~$p$ it becomes comparable to~$N^2$. This invalidates the
large-$N$ approximation for the sum. Closer analysis of our coherent
state shows that, due to the non-perturbative character of the initial
gauge configuration, the classical expectation values of the
operator~$\hat{O}_J$ in the coherent state are very large. The maximal
term in~\eqn{proba} is attained for large~$p$, which grows so fast
with~$N$ that one cannot neglect~$1/N^2$ and higher order corrections
in the denominator.  Moreover, summing all planar contributions does
not yield a good approximation either.

This makes the problem of calculating the energy distribution in the
outgoing state very hard to do analytically. In
section~\ref{s:U4results} we instead adopt a Monte-Carlo method in
order to compute the state norms, and subsequently evaluate the
sum~\eqn{proba} for all operators in the~U(4) case.  We show that, as
expected from string calculations, particles in the final state are
suppressed as their mass increases. We also show that, had one not
taken the \emph{full} norms in~\eqn{proba} into account, one would
incorrectly find that the energy distribution increases for more and
more massive particles. This is essentially due to the fact that
classical expectation values for all operators grow with their
dimension.
\medskip

Our results agree in a qualitative sense with results from previous
calculations on the gravity side. The calculations on the gravity side
have already been performed in the literature for decaying D-branes in
\emph{flat space}~\cite{Lambert:2003zr}. To compare these to the gauge
theory calculations, we ``embed'' these results
%for the spectrum of the
%decaying brane D-branes in the \emph{flat} space~\cite{Lambert:2003zr}
in the AdS space. A priori, there is no reason to expect that the flat
space results of the decay should be valid for branes in an AdS
background. However, since the D-particles in question are fully
localised in the bulk space, one expects that the flat space results
should carry over, at least when the radius of the AdS is
large.\footnote{To be precise, the D-particle dual to the sphaleron is
localised in the AdS part of space-time, while it is smeared on
the~$S^5$.}  There are two properties of the spectrum of the decaying
brane that we can compare with the dual gauge theory calculation. The
first property of the spectrum is constrained by the symmetries of the
system, and concerns emission amplitudes for the states on the leading
Regge trajectory. By slightly refining the calculation
of~\cite{Lambert:2003zr} in section~\ref{s:stringcalcs} we find that
all emission amplitudes for these states are zero. The same result is
then separately recovered on the gauge theory side by evaluating the
number operator for the corresponding dual composite operators.
% V3: footnote on smearing on S^5 added above.

More important is a second property of the spectrum, observed
in~\cite{Lambert:2003zr}, which reflects genuine dynamical features of
the decay. There is strong evidence~\cite{Sen:2002nu,Lambert:2003zr}
that the open strings decay \emph{fully} into closed string states,
i.e.~that there is no open string remnant left after the decay. This
conclusion is also supported by the matrix model calculations
of~\cite{McGreevy:2003kb}.  As shown in~\cite{Lambert:2003zr}, the
emission amplitudes are exponentially suppressed with the level of the
emitted string, at least for high levels (however, due to the
exponential growth of the available states, most of the energy of the
brane gets transferred into a high-density cloud of very massive
closed string states). By studying the dual gauge model we discover
the same qualitative feature: a \emph{suppression} of higher-mass
string states in the decay product.

\section{Decaying sphalerons in AdS/CFT}
\subsection{Classical instability of the sphaleron on $S^3\times \mathbb{R}$}

The first step in our analysis is to give a detailed description of
the decaying sphaleron on the gauge theory side.  Whereas the
sphaleron solution on $\mathbb{R}^4$ found by Klinkhamer and
Manton~\cite{Klinkhamer:1984di} is very complicated and not known
analytically, the situation is much simpler on
$S^3\times\mathbb{R}$. Not only is the solution known in this case,
but one can also find an analytic description of the classical decay
of this metastable state.

Following Drukker et al.~\cite{Drukker:2000wx}, one can get a sphaleron solution on $S^3 \times
\mathbb{R}$ by starting from the instanton solution on $\mathbb{R}^4$.
The latter is given by
\begin{equation}
\label{e:ansatz}
A_{\mu} = f(r) (\partial_{\mu}U) U^{\dagger}, \quad \quad
U=\frac{x^{\mu}\sigma_{\mu}}{r}, 
\quad \quad r^2 = x_0^2+x_i^2  \,,
\end{equation} 
where $f= r^2/(r^2 + a^2)$. This function interpolates between two
pure gauge configurations (i.e.~the two vacua) $f(r=0)=0$ and
$f(r=\infty)=1$. When $f(r)=1/2$, the system is at the top of the
potential barrier. By taking $f=1/2$ everywhere one gets a singular
solution to the equation of motion on $R^4$, which is the so-called
``meron''. The $f=1/2$ solution is, however, also a solution on
$S^3\times \mathbb{R}$, since this manifold can be conformally mapped
to $\mathbb{R}^4$ and Yang-Mills theory in four dimensions is
conformally invariant. The solution obtained in this way is the
Euclidean version of the ``sphaleron'', and is
non-singular.\footnote{The singularity of a meron originates from the
region $r \rightarrow 0$, since the action behaves as $S\sim
\int\!{\rm d}r\, r^{-1}$. After the conformal transformation the action density reduces
to a constant.} The Lorentzian version is the same, since the time
component of the potential of the sphaleron is zero; the solution is
completely time-independent.

We want to study the decay of the sphaleron, and we restrict to those
modes which preserve the \emph{spatial} homogeneity of the initial
sphaleron configuration. This is because we want to look at the decay
of the D-particle which sits at the ``origin'' of the anti-de-Sitter
space and is projected in the same way to all points on the
boundary. In other words, we only allow for time dependence, so that
energy-momentum tensor should be of the form
\begin{equation}
\label{e:Tform}
T_{00}= g(t)\,,  \quad \quad T_{ij} = h(t) g_{ij}\,,
\end{equation}
where $g_{ij}$ is the metric on $S^3$. So the ansatz we make is
\begin{equation}
\label{ansatz}
A = f(t)\, \Sigma^i \sigma_i\,,
\end{equation}
where $\Sigma^i$ are the three left-invariant one-forms. The energy
momentum tensor,
\begin{equation}
T_{\mu\nu} = \Tr( F_{\mu\rho}F_{\nu \sigma}g^{\rho\sigma}) - \frac{1}{4}g_{\mu\nu} \Tr( F^2 )
\end{equation}
reduces for our ansatz to the desired form~\eqn{e:Tform} with the
functions~$g$ and~$h$ given by
\begin{equation}
\begin{aligned}
g(t) &= - \frac{3}{2}R^2 \dot{f}^2 - 6 f^2(1-f)^2 \, , \\
h(t) &= - \frac{1}{2}R^2 \dot{f}^2 - 2 f^2 (1-f)^2 = - \frac{g(t)}{3}\,.
\end{aligned}
\end{equation}

\begin{figure}
\begin{center}
\includegraphics*[width=.45\textwidth]{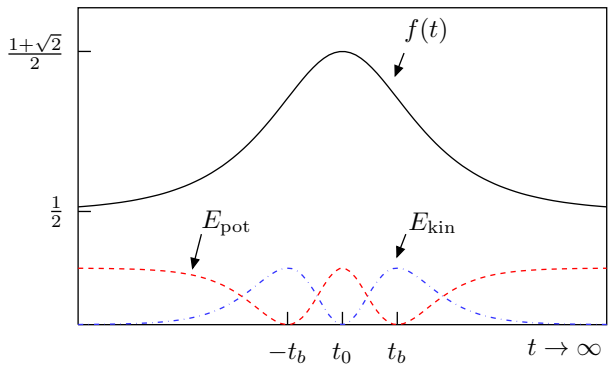}
% V2: f_- added to the figure.
\caption{The functions $f_\pm(t)$ of the decaying sphaleron on $S^3$ as
  given in~\eqn{e:fsol}, together with the kinetic and potential
  energy (with normalisation as given in~\eqn{e:Eformula} and $R=1$).}
\label{f:foft}
\end{center}
\end{figure}

To deduce what is the unknown function $f(t)$ we plug the ansatz into
the action and derive the action for this function. The value of the
action for our ansatz is
\begin{equation}
\label{e:reducedYM}
\begin{aligned}
S &= -\frac{1}{4 g_{\text{YM}}^2} \int\!{\rm d}t{\rm d}\Omega\, F_{\mu\nu}F^{\mu\nu}\\[1ex]
  &=  \frac{24\, \vol(S^3)}{4\, g_{\text{YM}}^2} \int\! \frac{{\rm d}t}{R} 
       \left(\frac{R^2}{2}\dot{f}^2 - 2 f^2 (1-f)^2\right) \, ,
\end{aligned}
\end{equation}
where $\vol(S^3)\equiv 2\pi^2$ denotes the volume of the unit sphere
and~$R$ is the radius of $S^3$ (also
see~\eqn{e:RS3metric},~\eqn{fs}). The equation of motion for the
function~$f$ is
\begin{equation}
\label{single}
R^2\,\ddot{f} + 4f(1-f)(1-2f) = 0 \, .
\end{equation}
This equation can be integrated once, yielding a conserved quantity,
namely the energy (i.e.~the component $T_{00}=48 \vol(S^3) E$)
\begin{equation}
E = R^2\,\dot{f}^2 + 4 f^2 (1-f)^2 \, .
\end{equation}
By introducing a new variable $H(t)$
\begin{equation}
f(t) = \frac{1}{2} (1 + H(t)) \, ,
\end{equation}
the expression for energy becomes
\begin{equation}
\label{e:Eformula}
4 E = R^2\,\dot{H}^2 + (1 - H^2)^2 \, , 
\end{equation}
which can be further integrated analytically
for~$E=\tfrac{1}{4}$. There are two solutions, corresponding to the
fact that the sphaleron can roll down on either side of the potential,
to the vacua with Chern-Simons number one and zero respectively. The
final result reads (see figure~\ref{f:foft})
% V2: second solution added.
\begin{equation}
\label{e:fsol}
f_\pm(t) = \frac{1}{2}\left(\frac{\pm\sqrt{2}}{\cosh\left(\frac{\sqrt{2}}{R} (t-t_0)\right)}
+ 1 \right)\, .
\end{equation}
One can check that these solutions are indeed solutions to the full
equations of motion, not just to the equations obtained from the
reduced action.  These solutions describe a configuration that starts
from the potential maximum at $t=-\infty$ (with zero velocity and
acceleration), rolls down one of the two sides of the hill and up the
other side, where it arrives at $t=t_0$. The minimum of the potential
energy~\eqn{e:Eformula} is reached when $H^2=1$ which corresponds to
\mbox{$t-t_0=\pm R\arccosh(\sqrt{2})/\sqrt{2}\approx \pm 0.62\,R$};
the evolution is symmetric around $t=t_0$.\footnote{After we had
derived this solution, we learned that it has been obtained before by
Gibbons and Steif~\cite{Gibbons:1994pq} and
Volkov~\cite{Volkov:1996es,Volkov:1996hm}, albeit in a different
context.}
% V2: reference to Gibbons & Steif added.

% V3: added paragraph on the periodicity
The periodicity of the whole process is natural from the AdS
perspective. Since AdS effectively acts as a box, the cloud of
outgoing radiation is refocused to the origin of the space, where it
arrives as fine-tuned radiation and ``re-builds'' the D-particle. In
this sense the D-particle never decays, since there is no real
dissipation of the energy in the system. However, in the limit of
large AdS radius, our flat-space intuition should (at least
approximately) hold. A natural point in time, which should be
associated to the decayed brane, is the point where the sphaleron has
rolled down to the the bottom of the potential, i.e.~when all
potential energy has been converted to kinetic energy.

The previous construction can easily be generalised to describe the
decay of a system of coincident D-particles.  The relevant sphaleron
configurations have been given by Drukker et
al.~\cite{Drukker:2000wx}. They are obtained by replacing the Pauli
matrices in~\eqn{e:ansatz} with Clifford algebra generators according
to
\begin{equation}
\label{e:sunsu2}
\sigma_\mu \rightarrow \gamma_\mu 
 = \begin{pmatrix}
   \sigma_\mu & 0 & \cdots & 0 \\
   0          & \sigma_\mu & \cdots & 0 \\
   \vdots     & \vdots & \ddots & 0 \\
   0          & 0      & \cdots & \sigma_\mu 
   \end{pmatrix}\, .
\end{equation}
This will make the various sphalerons sit in mutually commuting SU(2)
factors within U($N$).  In this case~\eqn{single} gets replaced by an
independent equation for each of the functions $f_i$, and the
solutions of those are given by~\eqn{e:fsol} which can differ only by
the value of~$t_0$. In what follows we will restrict to the situations
in which all these initial ``phases'' are the same, i.e.~in which all
D-particles start to decay at the same time. Since the field strengths
will also be block-diagonal, traces of powers of them will decompose
as sums of traces of the individual blocks.

\subsection{Coherent state description of the sphaleron decay}
\label{cohs}
In order to perform an analysis of the spectrum of the decay in the
gauge theory, as a first step one needs to construct a quantum state
describing the decayed sphaleron. For that purpose, it is useful to
think about the sphaleron decay in the following way.  Near the top of
the potential, most of the energy of the (perturbed) sphaleron
configuration is stored in the potential energy, which arises from the
non-linear terms in the Yang-Mills action. Precisely these
non-linearities in the action ensure the existence of the sphaleron
solution.  However, as the sphaleron decays, the potential energy of
the sphaleron gets transferred into kinetic energy, and near the
bottom of the potential valley all of the energy of the configuration
is stored in the kinetic energy. This can be seen most easily by
performing a finite gauge transformation (\ref{gtr}) on the
solution~(\ref{ansatz}) with gauge parameter
$\Lambda=U^{\dagger}$.\footnote{Alternatively, this can be seen from
the equation (\ref{single}) in which the coupling $g$ is restored; the
$g\rightarrow 0$ limit leads to the same equation of motion as
linearisation of $f$ around $f=1$.} The solution then reduces to
\begin{equation}
\label{e:Afull}
A_{\mu} = \tilde{f}(t)\, U^\dagger (\partial_\mu U)  \, , \quad
 \tilde{f} = f-1 \, .
\end{equation}
Near the bottom of the potential $\tilde{f}\approx 0$, which means
that the derivative part of the field strength, rather than the
non-linear (commutator) part, is dominant. The solution becomes
solution of the free Yang-Mills equations of motion on~$S^3 \times \mathbb{R}$
(written in the radiation gauge: $A_0 = \nabla_i A^i=0$)
\begin{equation}
\label{e:linearisedEOM}
\Big( -\partial_t^2 + \frac{1}{R^2} \big(\nabla_{S^3}^2 - 2\big) \Big) A^{\text{lin.}}_i =
0\, .
\end{equation}
Indeed, one can easily see that as~$t \rightarrow t_{\text{bottom}}$
the solution~\eqn{e:Afull} with $f$ given by~\eqn{e:fsol} is
very well approximated by the following solution of the linearised
equation of motion~\eqn{e:linearisedEOM}:
\begin{equation}
\label{e:Anearbottom}
A_i^{\text{lin.}} = 
-\frac{1}{4}\sin\left(\frac{2(t-t_{\text{bottom}})}{R}\right) \,
U^\dagger (\partial_i U) \,.
\end{equation}
Hence near the bottom of the valley, one can think about the
Yang-Mills configuration as dual to a coherent state of
non-interacting closed string states which are the product of the
D-particle decay.  Our goal will then be to determine numbers of
various (gravity) ``particles'' in this final coherent state. What we
precisely mean by this will be explained in the next section. Let
us first construct this coherent state.  

The fact that our solution abelianises near the bottom of the
potential valley allows us to write down a coherent state for this
configuration (see Cahill~\cite{Cahill:1974cs} and
Pottinger~\cite{Pottinger:1978ix} for related work). In order to do
so, we need an expansion of the free Yang-Mills gauge potential in
spherical vector harmonics. %
In the radiation gauge, an expansion is
given by (we refer the reader to Hamada and
Horata~\cite{Hamada:2003jc} for more on spherical harmonics)
\begin{multline}
\label{e:spherical}
A^{ab}_i =  \sum_{J,y,M} \Big(
  \hat{a}^{ab}_{JMy} \frac{e^{-(2J+1)\tau/R}}{\sqrt{2(2J+1)}} Y^i_{JMy}(\phi,\theta,\psi)\\[1ex]
 +\hat{a}^{\dagger ab}_{JMy} \frac{e^{(2J+1)\tau/R}}{\sqrt{2(2J+1)}} Y^{*i}_{JMy}(\phi,\theta,\psi)
 \Big)\,,
\end{multline}
where $M=(m,m')$ labels the representations of the simple factors of
SO(4)=SU(2)$\times$SU(2) and $y$ sums over the physical polarisation
states of the gauge field. We have also introduced matrix indices for
the adjoint representation of the U($N$) Lie-algebra. The other indices
run over the ranges
\begin{equation}
\label{indices}
\begin{aligned}
J &\geq \tfrac{1}{2}\,,\\
y &= \pm \tfrac{1}{2}\,,
\end{aligned}\quad
\begin{aligned}
-J-y &\leq\, m\,  &&\leq J+y\,,\\
-J+y &\leq\, m'\, &&\leq J-y\,.
\end{aligned}
\end{equation}
After quantisation, the operators $\hat{a}_{lmn}^\dagger$ and
$\hat{a}_{lmn}$ satisfy the canonical commutation relations
\begin{equation}
\label{e:canonical}
{}\Big[ \hat{a}^{ab}_{JMy}, \hat{a}^{\dagger cd}_{J'M'y'} \Big] 
  = g_{\text{YM}}^2\,
  \delta_{JJ'}\delta_{MM'}\delta_{yy'}\delta^{ad}\delta^{bc}\, .
\end{equation}

A coherent state (see Klauder and Skagerstam~\cite{b_klau1} for more
on coherent states and references to the literature) corresponding to
the classical configuration given by~\eqn{e:fsol} is constructed by
demanding that
\begin{equation}
\label{e:coherent_prop}
\hat{a}^{ab}_{JMy} |c\rangle = A^{ab}_{JMy}|c\rangle\, ,
\end{equation}
where $A_{JMy}$ are the coefficients appearing in the Fourier
decomposition of the classical sphal\-eron configuration, as
in~\eqn{e:spherical}. In the
Coulomb gauge $A_{0}^{ab} = 0$ the coherent state can be written as
\begin{equation}
\label{coherent}
|c\rangle = {\cal C}\, \exp\left( g_{\text{YM}}^{-2} \sum_{J,M,y} 
   \Tr\big(A_{JMy}\, \hat{a}^{\dagger}_{JMy}\big) \right) |0\rangle\,,
\end{equation}
The normalisation factor~${\cal C}$ is chosen such that $|c\rangle$ is of
unit norm and is given by
\begin{equation}
{\cal C} = \exp\Big( - \tfrac{1}{2}g_{\text{YM}}^{-2}
\Tr\big( \sum_{JMy} |A_{JMy}|^2\big)\Big) \,.
\end{equation}
A similar construction for the Klinkhamer-Manton sphaleron in
Yang-Mills-Higgs theory on $\mathbb{R}^{(3,1)}$ has been discussed
by Zadrozny~\cite{Zadrozny:1992yf}. The state~\eqn{coherent} is most natural
from the point of view of the gauge theory; we will discuss the
possibility of using alternative coherent states at the end of
section~\ref{s:AdSparticles}.

It is important to note that, due to the properties of the vector
spherical harmonics in~\eqn{e:spherical}, the coherent
state~\eqn{coherent} has been built from creation operators that
excite only physical excitations: the Gauss law constraint is
automatically implemented using these operators, since $\nabla \cdot
\hat{A}^{\text{free}} = 0$ holds as an operator equation. Hence the
coherent state~\eqn{coherent} is a legitimate state in the Hilbert
space of the free theory.

Nevertheless, the state~\eqn{coherent} does not yet provide a good
description of the system, as it is constructed using the oscillators
of the free theory and does not allow for a smooth deformation to the
interacting theory. At finite coupling, it does not satisfy the global
colour neutrality constraint. This constraint arises because the
commutator term in the Gauss law acts as a source, and by integrating
the constraint over the $S^3$ one finds that this total charge has to
vanish. One therefore imposes that the commutator part of the
non-abelian constraint vanishes also at zero
coupling~\cite{Sundborg:1999ue}. This constraint restricts the
coherent state to the colour singlet part,
\begin{equation}
\label{e:project}
|c_{\text{singlet}}\rangle = {\cal P}_{\text{singlet}} |c\rangle\,.
\end{equation}
In practise, however, we will neither write this projector nor
construct the projected state explicitly. This is because our
calculations always involve projections of the coherent state onto
states which themselves are colour singlets. Therefore the singlet
projection is imposed implicitly throughout. The only thing which we
have to keep in mind is that when the state~$|c \rangle$ is unit
normalised, the norm of~$|c_{\text{singlet}}\rangle$ is much smaller
than one; we will return to this issue in section~\ref{s:U4results}
when we discuss the decomposition of the coherent state in a specific
example.\footnote{We should also remark that in addition to the
configuration~\eqn{e:sunsu2} used in the construction of the coherent
state $|c\rangle$, there exists a whole family of inequivalent
configurations related to~\eqn{e:sunsu2} by large gauge
rotations. Since the parameters of this family do not have a
counterpart on the gravity side, one needs to integrate over these
configurations when calculating observables on both sides.  A similar
situation occurs when one calculates correlation functions in an~SU(2)
instanton background: while the size and position of the instanton do
have an AdS interpretation, the parameters describing the embedding of
the SU(2) instanton in SU($N$) do not, and thus have to be integrated
over.  For all the observables we will be calculating from
$|c\rangle$, these integrations lead to additional overall group
factors, which are irrelevant for our purposes. Hence in what follows,
we will ignore this technical subtlety.}

\subsection{Particles  in the AdS/CFT correspondence}
\label{s:AdSparticles}

In the AdS/CFT correspondence, we have a relation between string
states in the bulk and operators in the boundary. These operators are,
via the operator--state mapping, interpreted to create ``particles''
in the bulk theory at a particular point on the boundary. That is, one
needs to solve for the wave equation of the dual field in the bulk in
the presence of a delta source inserted at the boundary.  This means
that the states created in the bulk are not eigen momentum states, an
attribute which one usually associates to the notion of a particle in
field theories.  However, since the AdS/CFT correspondence is
formulated in position space rather than momentum space, these
definitions are natural in this context.  Our string calculation, on
the other hand, will be a flat space calculation, and for us it will
be more natural to use the standard notion of particles in the
bulk as angular momentum eigenstates. For that however, we will first
have to construct boundary operators that are dual to the bulk
angular momentum eigenstates.

The operator--state correspondence is usually discussed in the context
of radial quantisation of conformal field theories (see e.g.~Fubini et
al.~\cite{Fubini:1973mf} for a discussion in a four-dimensional
context). One first Wick rotates $\mathbb{R}\times \mathbb{R}^3$ to
the Euclidean regime and then performs a conformal transformation such
that the origin of $\mathbb{R}^4$ corresponds to $t=-\infty$ in the
original frame.  Operators inserted at the origin are then in
one-to-one correspondence with states in the Hilbert space. The entire
procedure can, however, be formulated without doing the conformal
rescaling, which is more natural in our setup since, as we have
discussed before, the gauge field configuration on $\mathbb{R}\times
S^3$ is non-singular while the one on $\mathbb{R}^4$ is singular.

The state corresponding to an operator with conformal weight~$w$ is
obtained by multiplying with the appropriate exponential of Euclidean
time and taking the limit $\tau\rightarrow-\infty$ (keeping only the
regular part): 
% V2: comment about taking the regular part added.
\begin{equation}
\label{e:Ostatedef}
\begin{aligned}
|\hat{O}^{(m)}_{\text{weight-}w}\rangle &= \lim_{\tau\rightarrow-\infty} 
  \,\Big\{e^{-w\tau}
  \hat{O}^{(m)}_{\text{weight-}w}(\tau)\Big\}
  \big|0\big\rangle\\[1ex]
&\equiv \hat{O}^{(m)}_{\text{weight-}w} |0\rangle\, .
\end{aligned}
\end{equation}
The last expression shows the shorthand notation that we will use in
order not to clutter expressions unnecessarily.  The hermitian
conjugate of an operator is given by
\begin{equation}
\Big(\hat{O}(\tau)\Big)^\dagger
 = \hat{O}^\dagger(-\tau)\, .
\end{equation}
This procedure mimics the operator--state mapping on $\mathbb{R}^4$
but avoids technical problems related to solutions which become
singular after the conformal transformation.

The operators which we use in~\eqn{e:Ostatedef} are independent of the
angular coordinates on the sphere, i.e.~they are obtained from the
position dependent operators as follows
% V2: removed 'time independent' above, added \tau below and 
%     combined the angular coordinates into \phi_i.
\begin{equation}
\label{e:pola}
\hat{O}_{w}^{(m)}(\tau) = K^{(m)}_{w} \int_{S^3}\! {\rm d}\Omega\,
\, \hat{O}_{w}^{\mu_1...\mu_s}(\tau,\phi_i)\, Y_{\mu_1...\mu_s}^{(m)}(\phi_i) \, .
\end{equation}
Here $Y^{(m)}$ denote the lowest lying tensor spherical harmonics for
a given spin~$s$. The index~$m$ labels the degeneracy of such
harmonics. The normalisation constants~$K^{(m)}_w$ are chosen such
that the states constructed using~\eqn{e:Ostatedef} are of unit norm.
Note that the multiplication with the time dependent exponent
in~\eqn{e:Ostatedef} selects out composite operators of the required
conformal dimension, but when one expresses these operators in terms
of elementary creation and annihilation operators, one explicitly sees
that different operators~$\hat{O}$ are not orthogonal.  It is only
after the integration~\eqn{e:pola} that one obtains a set of
orthogonal states. See appendix~\ref{a:s2opstate} for an explicit
example on~$S^2$.

Multi-particle states are obtained by acting repeatedly with
the~$\hat{O}_w$ operators on the vacuum, in analogy with normal
creation operators for elementary particles. In contrast to elementary
operators, however, there is no simple number operator which can be
used to count the number of composite excitations in a given state. It
is true that
\begin{equation}
\label{e:OdaggerO}
{} [\hat{O}, \hat{O}^\dagger ] = 1 + {\cal O}(N^{-2}) \,,
\end{equation}
and one might expect that this leads to a well-defined number
operator~$\hat{O}^\dagger \hat{O}$. However, the coefficients that multiply
the~$1/N^2$ corrections in~\eqn{e:OdaggerO} are operators, not
c-numbers. As a consequence, the strength of the~$1/N^2$ corrections
depends on the state in which the number operator is evaluated,
\begin{equation}
\label{noco}
\langle n | \hat{O}^\dagger \hat{O} | n\rangle = n + \sum_i \frac{c_i(n)}{N^{2i}}\,.
\end{equation}
The numbers $c_i(n)$ can become arbitrarily large
when~$n\rightarrow\infty$. Since the coherent state contains such
highly excited states, the operator $\hat{O}^\dagger \hat{O}$ cannot
be used as a number operator, not even in the $N\rightarrow\infty$
limit.\footnote{An proper number operator for composite particles,
which produces the exact occupation number rather than an expression
which is only correct up to $N^{-2}$ corrections, has been constructed
by Brittin and Sakakura~\cite{Brittin:1980ev,brit2}. However, their operator is very
complicated and difficult to handle in practise. We prefer to follow a
different route here.} We will encounter an explicit manifestation of
this fact in section~\ref{s:U4results}, see in particular
figure~\ref{f:normcomparison}.

We will therefore follow a different route. Instead of trying to use a
number counting operator applied to the coherent state, we will simply
project the coherent state on each state in the Hilbert space of
multi-particle states. Subsequently, using these probabilities, we
will calculate the average energies and particle numbers.  Details of
this procedure will be discussed in the section~\ref{s:U4results}.

Let us end this section with a comment on alternatives to the coherent
state~\eqn{coherent}. From the point of view of the dual string
theory, it might seem more natural to construct a coherent state using
the composite operators~$\hat{O}^\dagger_J$ in the exponent, rather
than the elementary ones~$\hat{a}^\dagger$. After all, the~$\hat{O}_J$
correspond to elementary string excitations. However, a state of
the form
\begin{equation}
|\tilde c\rangle = \tilde{{\cal C}} \exp\Big(
  \sum_i O^{\text{class.}}_i\,\hat{O}^\dagger_i
\Big)|0\rangle
\end{equation}
is not a coherent state in the sense of~\eqn{e:coherent_prop}. The
expectation value of an operator in this coherent state does not equal
the classical value of that operator,
\begin{equation}
\big\langle \tilde c\big|\, \hat{O}_i\, \big|\tilde c\big\rangle \not= O^{\text{class.}}_i\,,
\end{equation}
not even up to $1/N$ corrections. The reason for this is essentially
given in equation~\eqn{noco}, with~$|n \rangle$ now being given
by~$|n\rangle = \big(\hat{O}^\dagger_i\big)^n\,|0 \rangle$.  This is
our prime motivation to use~\eqn{coherent} as the sphaleron coherent
state.

\section{The decay spectrum}
\subsection{Counting procedure and symmetry considerations}
\label{s:particles}
Having constructed a coherent state in the gauge theory which is dual
to the final state of the D-particle decay~(see section~\ref{cohs}),
we now want to extract information from it about particle numbers in
the decay product. By particle counting, we mean counting of the
states constructed in the previous section.  The main subtlety for
this calculation lies, as we have discussed in the previous section,
in the fact that we want to count states created by composite rather
than elementary operators.  In this section we will outline the
general procedure which we will use to calculate these numbers, and
then apply it to a special class of operators whose behaviour seems
to be fully determined by the symmetries of the problem.

The basic ingredient in our particle counting procedure is the
probability to find a particular multi-particle state of composite
particles in the coherent state. The probability of finding a
multi-particle state consisting of $p_1$ particles of type $O_{J_1}$,
$p_2$ particles of type $O_{J_2}$ etc., is given by
% V2: rewrote states to avoid issues concerning state
% orthonormalisation.
\begin{multline}
\label{expectations}
{\cal P}(p_1;p_2;\ldots;p_M) :=\\[1ex]
     \frac{\left|\, \Big\langle 
     (\hat{O}_{J_1})^{p_1} \ldots (\hat{O}_{J_M})^{p_M}\,
\Big|\,c\Big\rangle \, \right|^2}{\Big\langle \big( \hat{O}_{J_1} \big)^{p_1} \ldots 
            \big( \hat{O}_{J_M} \big)^{p_M}\, \Big|\,
            \big( \hat{O}_{J_1} \big)^{p_1} \ldots 
            \big( \hat{O}_{J_M} \big)^{p_M} \Big\rangle\,\big\langle
     c\big|c\big\rangle}\,.
\end{multline}
For this to work it is of course crucial that the basis of
multi-particle states is constructed to be orthogonal.
% V2: footnote added on orthogonalisation of states
\footnote{\label{f:ortho} Even if one has two orthogonal states $O_1^\dagger
  |0\rangle$ and $O_2^\dagger|0\rangle$ created using composite
  creation operators, it is generically not true that \mbox{$\langle 0
	 | (O_1)^n (O^\dagger_2)^n |0\rangle=0$}. The notation used
  in~\eqn{expectations} is meant to indicate that proper subtraction
  terms are included, such as to make all states appearing in the sum
  orthogonal. Using these orthogonal states, the projection operator
  appearing in~\eqn{e:project} takes the form
\begin{multline}
{\cal P}_{\text{singlet}} = 1 + \sum_{\{p_1,p_2,\ldots p_n\}}
1/{\cal N}_{p_1,p_2,\cdots p_n}\\[1ex] \times\big|(O_1)^{p_1}
(O_2)^{p_2} \ldots (O_n)^{p_n} \big\rangle\big\langle (O_1)^{p_1}
(O_2)^{p_2} \ldots (O_n)^{p_n}\big|  \,,
\end{multline}
where the ${\cal N}_{p_1,p_2,\ldots p_n}$ are the norms of the
states.} By definition, the average number of particles of the
type~$\hat{O}_{J_i}$ present in the coherent state is now given by
\begin{align}
\label{e:numbers}
N(J_i) &:= 
  \sum_{p_1=0}^\infty \cdots \sum_{p_M=0}^\infty
     p_i\, {\cal P} (p_1;p_2;\ldots;p_M)\,. 
\intertext{The energy stored in these particles, as measured with
respect to the global time in the bulk, is given by the conformal
dimension of the corresponding operators. Therefore, the total energy
  is given by the expression}
\label{e:energies}
E(J_i) &:= 
  \sum_{p_1=0}^\infty \cdots \sum_{p_M=0}^\infty
     \Delta_{J_i}\, p_i\, {\cal P} (p_1;p_2;\ldots;p_M)  \, , 
\end{align}
where $\Delta_{J_i}$ is the conformal dimension of the operator
$\hat{O}_{J_i}$. This is actually why we interpret $\hat{O}^\dagger\cdots
\hat{O}^\dagger|0\rangle$ as a multi-particle state: the supergravity energy
is simply the sum of the constituent energies, despite the fact that
the norm does not factorise as the product of individual particle
norms.  

Due to the general properties of the coherent state, an
evaluation of the numerators in~\eqn{expectations} is
straightforward. It amounts to evaluating the classical expressions
for the (multi-)particle operators using the positive frequency
part~$A^+$ of the rolling sphaleron solution, near the bottom of the
potential.  When doing this calculation one also needs to use
formula~\eqn{e:pola}; i.e. for each particles in the state separately,
one needs to remove the~$e^{\tau}$ factors and then project onto the
corresponding lowest-lying harmonics.  Hence, even though we know the
full time dependent sphaleron solution~\eqn{ansatz}--\eqn{e:fsol}, we
need only the part of the solution at the end of the decay for the
evaluation of~\eqn{expectations}.

Since we are looking at a very simple, spherically symmetric decay,
the final phase of the decay is very much constrained and is basically
independent of the shape of the potential: it is given by an S-wave on
the three-sphere, with an amplitude determined by the height of the
potential. We have given this solution in~\eqn{e:Anearbottom}, and the
Euclidean version of its \emph{positive-frequency} part is given by
\begin{equation}
A_i^+ = \frac{i}{8} \exp\Big( \frac{2}{R}
  \big(t-t_{\text{bottom}}\big)\Big) U^\dagger (\partial_i U) \equiv f^+(t) U^\dagger (\partial_i U) \,  .
\end{equation}

As we will see in the next section, the real problem in calculating
the numbers of different states is related to the calculation of the
denominators in~\eqn{expectations}.  However, for the class of
operators which are absent from the decay spectrum, i.e.~for which
numerators in~(\ref{expectations}) vanish, one does not need to worry
about this issue.  The first operator in this class is the energy
momentum tensor.  Its vanishing implies that there is no gravitational
radiation in the bulk, a feature which is expected from the symmetry
of the problem. Namely, since the decay is spherically symmetric, there
are no quadrupole moments turned on, and hence no gravitational
radiation can be produced.\footnote{Note that the expression which
vanishes is the energy momentum tensor evaluated on the positive
frequency part of the solution:~$|\langle 0 |\hat{T}_{\mu\nu}|c
\rangle|^2 = |T_{\mu\nu}(A^+_{\text{coherent}})|^2=0$. On the other
hand, the classical expression for the energy momentum tensor of the
full configuration is non-zero:~$T_{\mu\nu}(A^+ + A^-) \neq 0$. Note
also that the~$\hat{T}_{\mu\nu}$ which is used here is the
\emph{abelian} expression for the energy momentum tensor, since all
our calculations are done in the free theory. It would be interesting
to extend the above analysis to include interactions. In that case,
however, there may be non-trivial corrections to the coherent state,
and both gauge bosons and scalars will contribute to the numerators
in~\eqn{expectations}.}

The lowest-mass SO(6) singlet that arises from the $S^5$ reduction of
the NS-NS two form is given by~\cite{Das:1998ei,Kim:1985ez}
\begin{equation}
{\cal O}_{\mu\nu} \sim \Tr \left( \frac{1}{2} F_{\nu \alpha} F^{\alpha
  \beta} F_{\beta \mu} 
+ \frac{1}{8} F_{\alpha \beta} F^{\alpha \beta} F_{\mu\nu} \right) \, . 
\end{equation}
The associated state also has a vanishing overlap with the sphaleron
coherent state.  This is due to the fact that this operator is cubic
in the field strength, and our gauge potentials are
``abelianised''~SU($N$) fields, as explained in section~\ref{cohs}.

% V2: removed 'surprisingly'
In the massive string sector, we also find that the radiation
associated to all twist-two fields vanishes. Namely,
\begin{equation}
N\big(\hat{O}_{\mu_1\cdots \mu_s}\big) = 0\,, \quad \text{for all
  $s$}\, , 
\end{equation}
where the operators are given by
\begin{multline}
\hat{\cal O}_{\mu_1 \cdots \mu_s} = \\[1ex]
\frac{\vol(S^3)^{-1} \, R^{2+s}}{\sqrt{2(g_{YM}^2 N)^{2}}}
  \norder{\Tr \big(F_{\nu(\mu_1}\nabla_{\mu_2}...\nabla_{\mu_{s-1}}
  F_{\mu_s)}{}^{\nu}\big)} \\[1ex] -\, (\text{traces}) \, .
\end{multline}
Here the $s=2$ operator corresponds to the graviton. We will see in
section~\ref{s:stringcalcs} that all these results can be matched with
the string theory prediction.

We believe that the vanishing of the amplitudes of the twist-two
operators is related to the symmetries of the system, rather than to
genuine dynamical properties. Hence, in order to gain insight in real
dynamics of the problem, we need to consider number operators for 
generic states, which we will do in section~\ref{s:U4results}.

\subsection{Expectations from the string side}
\label{s:stringcalcs}
 
The lack of knowledge about string quantisation on the $AdS_5\times
S^5$ background makes a direct study of D-particle decay in this
background impossible.  However, because the D-particle is a fully
localised object, one expects that its static and dynamic properties
are, at least for large radii, similar to those of the D-particle in
flat space~\cite{Drukker:2000wx}. We will employ this argument to use
a flat-space string calculation, rather than one in AdS, when making a
comparison to the gauge theory results.

In order to analyse the decay products of an unstable D-brane in
string theory, one has to solve for the closed-string
field~$|\Psi_c\rangle$ in the presence of a time-dependent brane
source,
\begin{equation}
\label{e:sftwithB}
(Q+\bar Q)|\Psi_c\rangle = |B\rangle\,.
\end{equation}
Here $|B\rangle$ is the boundary state for the unstable D-brane while
$Q$ and $\bar{Q}$ are the BRS operators.  The solution for
$|\Psi\rangle$ as well as the late-time behaviour was analysed
in~\cite{Lambert:2003zr,Sen:2004zm}. For the final state of the
decaying D-particle it takes the simple form
%\begin{widetext}
\begin{equation}
\label{e:Psifinal}
\begin{aligned}
|\Psi_c^\infty\rangle &:= \lim_{t\rightarrow \infty} |\Psi_c\rangle
 \propto\\
 \int\!&{\rm d}^{25}k_{\perp}\, 
  \sum_{L\geq 0} \exp\Big[ \sum_{n=1}^\infty
  -\frac{1}{n} \alpha^0_{-n} \bar\alpha^0_{-n} 
  + \alpha_{-n}^i \alpha^i_{-n} \\
&\hskip.2\textwidth - \text{ghosts} \Big] 
  \Big|_{\text{level $L$}} \\[1ex]
\times&\Big(  f(L,k_{\perp}) \big| k^0 = \omega_k, k_{\perp}, k_{\parallel}=0 \big\rangle\\
     &\quad\quad+ f^*(L,k_{\perp}) \big| k^0 = -\omega_k, k_{\perp}, k_{\parallel}=0 \big\rangle\Big)\,.
\end{aligned}
\end{equation}
%\end{widetext}
Here~$L$ denotes the oscillator level and $f(L,k_\perp)$ is a function
dependent on the level and the transverse momentum.\footnote{There is
a subtlety concerning terms in the outgoing state which grow
exponentially in time. Their interpretation is at present not entirely
clear~\cite{Sen:2004zm}. Moreover, there exists an alternative
derivation in which such terms are not
present~\cite{Gutperle:2003xf}. We do not want to go into a discussion
of this issue here, and consider only the terms which are finite at
late times.} This final state can now be projected onto on-shell
closed string states.

The twist-two states which we are interested in are associated with
vertex operators which, for a given level, carry maximal spin. This is
achieved by using the maximal number of creation operators for fixed
level, i.e.~by building the state using only $\alpha_{-1}^{\mu}$
operators. More precisely, the particle numbers in the final state
will be determined from the following overlaps
% V2: changed '-1' to '1'.
\begin{equation}
\label{e:theamplitude}
S^{\mu_1\ldots\mu_n \nu_1\ldots\nu_n} 
  = \big\langle k^{\mu} \big| \,\alpha_{1}^{\mu_1}\ldots\alpha_{1}^{\mu_n}\,
                \bar\alpha_{1}^{\nu_1}\ldots\bar{\alpha}_{1}^{\nu_n}\,
    \big| \Psi^\infty_c \big\rangle\, ,
\end{equation}
where $k^\mu$ is the momentum of the centre of mass of the closed
string, related to the level~$n$ via the mass shell condition $k^2 =
2(1-n)$.  The twist-two states are associated with the part of these
vertex operators which have maximal spin, that is, they are built out
from the vertex operators by contracting them with polarisation
tensors satisfying
\begin{equation}
\label{e:polconds1}
\epsilon_{\mu_1\ldots\mu_n\nu_1\cdots\nu_n} = 
  \epsilon_{(\mu_1\ldots\mu_n\nu_1\cdots\nu_n)}\,,\quad
\eta^{\mu_1\nu_1} 
  \epsilon_{\mu_1\ldots\mu_n\nu_1\cdots\nu_n} = 0\, .
\end{equation}
For these polarisation tensors, it is easy to see that the
projections~\eqn{e:theamplitude} vanish. Since only the~$n=1$
oscillators appear in the twist-two states, the exponent
of~\eqn{e:Psifinal} effectively reduces to~$\eta_{\mu\nu}
\alpha^\mu_{-1}\bar{\alpha}^\nu_{-1}$.  All
projections~\eqn{e:theamplitude} then become proportional to traces of
the polarisation tensors, which vanish by~\eqn{e:polconds1}. There is
therefore no twist-two radiation; in particular, there is no
gravitational radiation (which is expected because there is no
quadrupole moment). The radiation into NS-NS two-form states also
vanishes, because the polarisation tensor is in this case anti-symmetric.

All of these considerations change when one does not select the
highest-spin state from the polarisation tensor (i.e.~when one does
not impose~\eqn{e:polconds1}). In particular, the dilaton radiation
will not vanish. These observations match the calculations on the
Yang-Mills side performed in the previous section.\footnote{The
conclusions also rely crucially on the fact that we are restricting
here to the D-particle case. The twist-two radiation is, for other
boundary states, generically no longer zero. The final
state~\eqn{e:Psifinal} will be more complicated, and the $n=1$ part of
the exponent will not reduce to
$\eta_{\mu\nu}\alpha^\mu_{-1}\bar{\alpha}^{\nu}_{-1}$.}

\subsection{Decay products in U(4) -- dynamical considerations}
\label{s:U4results}

For a generic operator, the calculation of the numerators
in~\eqn{expectations} is the same as in section~\ref{s:particles}, and
amounts to evaluating the classical expression of the (abelianised)
operator using the positive frequency part of the decayed solution.
The main technical problem arises when evaluating the denominators
of~\eqn{expectations}. To illustrate this, let us consider a
``simplified'' model, based on a non-abelian scalar field. This model
exhibits all of the technical subtleties associated to the
determination of the decay products. The crucial ingredients of the
vector coherent state, namely that it is constructed from the
lowest-lying spherical harmonics and that it depends
non-perturbatively on the coupling constant, are preserved by this toy
model. However, it avoids the inessential technical complications
associated to the evaluation of tensor spherical harmonics in the
numerators of~\eqn{expectations}.

The coherent state for a given classical configuration in this
non-abelian scalar theory is given by
\begin{equation}
\label{e:simplecoh}
\begin{aligned}
|c\rangle &= {\cal C} \exp\left( \frac{1}{g_{\text{YM}}^2} \Tr\big( a\,
\hat{a}^\dagger\big) \right) |0\rangle\,, \\[1ex]
{\cal C} &= \exp\left(
-\frac{1}{g_{\text{YM}}^2} \Tr\big( a^\dagger a\big)\right)\,.
\end{aligned}
\end{equation}
This mimics the construction~\eqn{coherent}. The unit normalised (at
leading order in $1/N$ expansion), single-trace operators which create
particles in the out vacuum are
\begin{equation}
\label{sample}
\hat{O}_J^\dagger = \frac{1}{\sqrt{\strut J (g_{YM}^2 N)^J}} \Tr\big( (\hat{a}^\dagger)^J \big) \, .
\end{equation}
These operators are coordinate independent operators, obtained using a
procedure similar to~\eqn{e:pola}.

With the above normalisation of the
operator, the numerators and hence probabilities in~\eqn{expectations} depend on the
Yang-Mills coupling in a non-perturbative fashion,
\begin{equation}
\label{argu}
\Big|\langle 0 | \big(\hat{O}_J\big)^p |c \rangle\Big|^2 = {\cal C}^2
				 \left| \frac{ \Tr\big((a^+)^J\big)}{\sqrt{\strut J (
				 g_{YM}^2 N)^J}} \right|^{2p} \, \equiv \frac{{\cal C}^2}{
				 J^p} \, \left({\frac{\eta_J^{2}}{\lambda^{J}}} \right)^p
				 \, ,
\end{equation}
% V2: refined statement about the behaviour of $\eta_J$ as a function
%     of the number of D-particles.
(where the last equality defines $\eta_J$; note that it is of the
order~$N$ for the configuration~\eqn{e:sunsu2} and generically scales
as the number of D-particles).  This reflects the fact that our
original sphaleron configuration is a non-perturbative solution of the
equations of motion. Note also that the only way in which the
coupling~$\lambda$ appears in~\eqn{e:numbers} and~\eqn{e:energies} is
through the combination~$\eta_J^2/ \lambda^J$.

\begin{figure*}
\begin{center}
\includegraphics*[width=.9\textwidth]{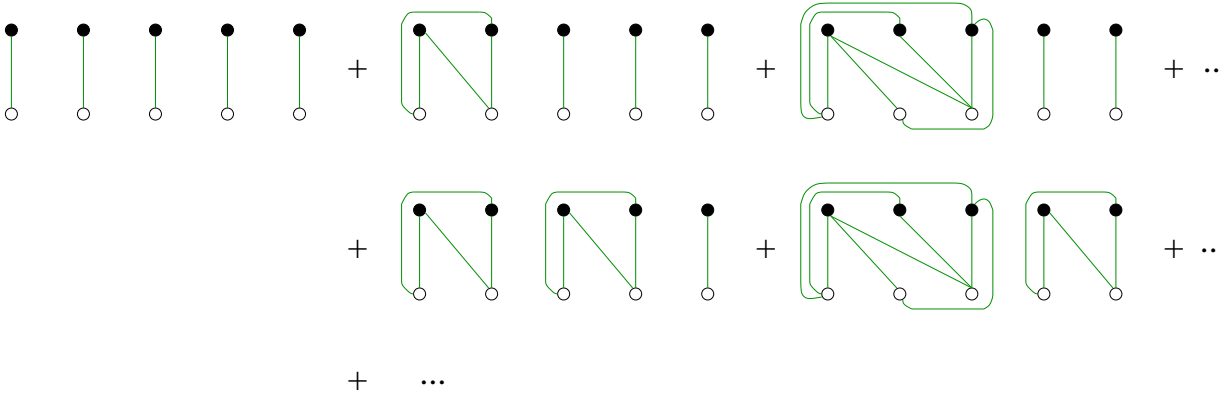}
\caption{Generic graphical expansion of the planar part of the norms
  required in~\eqn{expansion}. For simplicity we have only depicted
  the case in which there is only one type of operator; open dots
  represent $\Tr(\hat{a}^J)$ for fixed~$J$ and black dots their
  hermitian conjugates. The lines represent planar multiple
  contractions of elementary oscillators. The graphs displayed here
  correspond to typical ``large'' terms in each of the expressions
  in the sum~\eqn{expansion}.}
\label{f:fourblobs}
\end{center}
\end{figure*}

The complicated part of the calculation of the average particle
numbers and energies is the computation of the norms for the states
with an arbitrary number of particles.  The norm of the state with
$p$~identical particles can be written as (see also
figure~\ref{f:fourblobs})
%\begin{widetext}
\begin{equation}
\label{expansion}
\begin{aligned}
 &\big\langle (\hat{O}_J)^p\, (\hat{O}_J^\dagger)^p \big\rangle = 
   p!\, \big\langle (\hat{O}_J)\, (\hat{O}_J^\dagger)\big\rangle^p 
 \\[1ex]
&+ \binom{p}{2}^2 \big\langle
   (\hat{O}_J)^2\,(\hat{O}_J^{\dagger})^2\big \rangle_{\text{conn.}} 
   (p-2)! \big\langle (\hat{O}_J)\, (\hat{O}_J^\dagger) \big\rangle^{(p-2)} \\ 
& +\binom{p}{3}^2 \langle  \hat{O}_J^3  \hat{O}_J^{\dagger 3} \, \rangle_{\text{conn.}} (p-3)! \langle
\hat{O}_J \hat{O}_J^\dagger \rangle^{(p-3)} \\[1ex]
&+ \binom{p}{2}^2 \binom{p-2}{2}^2  \big\langle  (\hat{O}_J)^2
   (\hat{O}_J^{\dagger})^2  \big\rangle_{\text{conn.}}^2  \frac{(p-4)!}{2!} \langle
\hat{O}_J \hat{O}_J^\dagger \rangle^{(p-4)} \\[1ex] &+ ...
\end{aligned}
\end{equation}
%\end{widetext}
The first term is at a leading order independent of $1/N$, the second
is suppressed as~$1/N^2$, the last two terms both scale as~$1/N^4$, and
so on.  A similar but more complicated expansion can be written for
states involving more than one type of particle.

Naively, one might expect that in the large-$N$ limit, all but the
leading term~$p!$ in this expansion can be omitted. In
formula~\eqn{e:numbers}, this would produce an exponential dependence
on the expectation values for the operators $\hat{O}_J$.  Since the
arguments of the exponent~\eqn{argu} \emph{increase} with conformal
dimension~$J$, one would conclude that the number of particles
produced during the decay \emph{increases} with the mass of the
particle.  However, this kind of truncation of (\ref{expansion}) does
not make sense in the case of the \emph{non-perturbative} coherent
state~\eqn{e:simplecoh}, as it would actually produce
probabilities~\eqn{expectations} which are larger than one. The point
is that since the numerator~\eqn{argu} is very large, the maximal
probabilities are attained for large values~$p^{\text{max}}$
of~$p$. Moreover,~$p^{\text{max}}$ grows with~$N$, hence in the
large-$N$ limit the sub-leading terms in~\eqn{expansion} become more
and more relevant, and are actually \emph{comparable} to the leading
term.

In trying to estimate how fast the norms (\ref{expansion}) have to
grow with~$p$, one can see that even an exponential growth of the
norms, say as~$p!\, \gamma^p$ ($\gamma=\text{const}.$), does not lead
to reasonable results.  Namely, if we consider the
expression~$\sum_{p} {\cal P}(J,p)$, which has to be smaller than one,
and assume exponential growth of norms, we would find that this sum
behaves as
\begin{equation}
\label{e:pos-prob}
\begin{aligned}
\sum_{p=0}^\infty {\cal P}(J,p) &= {\cal C}^2 \sum_{p=0}^\infty
\frac{1}{p!} \left(\frac{\eta_J^2}{\lambda^J \gamma}\right)^p \\[1ex] 
&= \exp\left(\frac{\eta_J^2}{\lambda^J \gamma}\right) \exp\left(- \frac{N}{\lambda}
\Tr(a^\dagger a)\right)\,.
\end{aligned}
\end{equation}
Hence we see that even when $N\rightarrow\infty$ (while keeping
$\lambda$ arbitrary but smaller than one) the result will always be
larger than~1 for some value of~$J$. Since the calculation of the
average number of particles requires a summation over all~$J$, we
conclude that we cannot assume this behaviour of the norms.\footnote{Note
that if we would have had a perturbative coherent state instead of a
\emph{non-perturbative} one, the classical expectation values~$a$ in
% V2: removed '^2' on g_{YM}.
(\ref{e:simplecoh}) would be of the form $a = g_{YM} \eta$, with
$\eta$ a number independent of the coupling constant.  Hence formula
(\ref{e:pos-prob}) would be replaced with
$$
\label{e:pert-prob}
\sum_{p=0}^\infty {\cal P}(J,p) = {\cal C}^2 \sum_{p=0}^\infty
\frac{1}{p!} \left(\frac{\eta_J^2}{N^J \gamma}\right)^p =
\exp\left(\frac{\eta_J^2}{N^J \gamma}\right) \exp\left(- \Tr(a^\dagger a)\right)\,.
$$
We now see that a truncation to the first term in~\eqn{expansion}
(i.e.~setting $\gamma=1$) produces reasonable results for the
probabilities~\eqn{expectations}.}

The situation which we face here is similar in spirit to the
double-scaling BMN limit. As observed by Kristjansen et
al.~\cite{Kristjansen:2002bb} and Constable et
al.~\cite{Constable:2002vq}, in the limit $N \sim J^2
\rightarrow\infty$ correlators in general receive contributions from
non-planar graphs of all genera.  In this case, a new expansion
parameter~$J^2/N$ appears. In our case, $N\rightarrow\infty$ as well,
but now the additional parameter which becomes large is the value of
the~$p_i$ for which the sum~\eqn{e:energies} has its maximum term.  It
would be interesting to understand whether our system also exhibits a
double-scaling limit in which some ratio of powers of~$p$ and~$N$ is
kept fixed.

In order to determine the correct values of the norms of the states,
it is useful to write the norms in terms of correlators of a complex
matrix model,
%\begin{widetext}
\begin{multline}
\label{e:mcintegral}
\big\langle 0 \big| \Big[ \big( \hat{O}_{J_1} \big)^{p_1} \ldots 
            \big( \hat{O}_{J_n} \big)^{p_n}\Big] \, \Big[
            \big( \hat{O}^\dagger_{J_n} \big)^{p_n} \ldots 
            \big( \hat{O}^\dagger_{J_1} \big)^{p_1}\Big] \, \big|0
            \big\rangle=\\[1ex]
 \int\!{\rm d}A{\rm d}\bar A\, \Big[\big( {O}_{J_1} \big)^{p_1} \ldots 
            \big( {O}_{J_n} \big)^{p_n}\Big] \, \Big[
            \big( {O}^\dagger_{J_n} \big)^{p_n} \ldots 
            \big( {O}^\dagger_{J_1} \big)^{p_1}\Big] \\[1ex]
\times \exp\Big( - \Tr( A^\dagger A ) \Big)\,,
\end{multline}
%\end{widetext}
The measure used here is simply a separate integral over the real and
imaginary parts of the complex matrix $A$, normalised to give unit
result when all~$p_i$ in the expression above are zero,
\begin{equation}
\int\!{\rm d}A{\rm d}\bar A = \pi^{-N}\prod_{a,b=1}^N {\rm d}(\Real
  A_{ab})\,\,\, {\rm d}(\Imag A_{ab})\,.
\end{equation}
This approach has been used by~Kristjansen et
al.~\cite{Kristjansen:2002bb,Beisert:2002bb} in order to compute
several special cases of~\eqn{e:mcintegral} analytically. It is still
an open problem to extend those exact results to the entire class of
correlators, in particular to general situations for
which~$p_i>2$. Because we will need these very general correlators, we
have decided to use an alternative approach, in which the integral is
evaluated using Monte-Carlo methods. This provides us with a
technically straightforward way to extract the norms for arbitrary
operator insertions, even for very large~$p_i$. Our results will, for
this reason, of course be restricted to a fixed value for~$N$, and
computer resources put a practical limit on the maximum value that can
be handled. Nevertheless, we will see that interesting results can be
obtained this way.

\begin{figure}
\begin{center}
\includegraphics*[width=.45\textwidth]{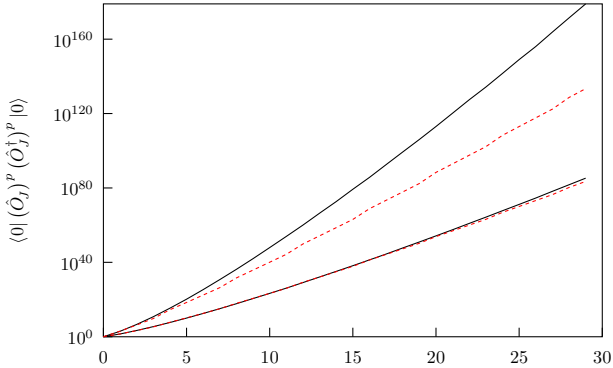}
\caption{Norms of states built from $J=2$ operators (lower two curves)
  and $J=4$ operators (upper two curves), as a function of the total
  number of operator insertions, for~U(4). The dashed lines are the
  estimates based on the first two columns of graphs in
  figure~\ref{f:fourblobs}.  The continuous lines are the complete
  norms extracted from the Monte-Carlo analysis.}
\label{f:normcomparison}
\end{center}
\end{figure}
Before we discuss the results, let us present numerical evidence which
illustrates the necessity of taking the~\emph{full norms}
in~\eqn{expectations} into account, i.e.~all planar and non-planar
contributions. We compare the results obtained by summing up a large
class of planar diagrams in~\eqn{expansion} with the numerical results
obtained using the Monte-Carlo integration.  For the U(4) case, the
Monte-Carlo results are depicted in figure~\ref{f:normcomparison} and
compared to the analytic answer obtained by restricting to the first
two columns of graphs in figure~\ref{f:fourblobs}; these columns
contain graphs with an arbitrary number of connected four-blob
elements. Clearly, the full norms deviate substantially from this
estimate. Hence, in the remainder we will only employ the norms
obtained by Monte-Carlo integration of~\eqn{e:mcintegral}. Note also
from figure~\ref{f:fourblobs} that for large $p$, the deviation from
the exact norm grows with increasing $J$ (and it also grows with
increasing~$N$), and does not improve as one might naively expect.

\begin{figure}
\begin{center}
\includegraphics*[width=.45\textwidth]{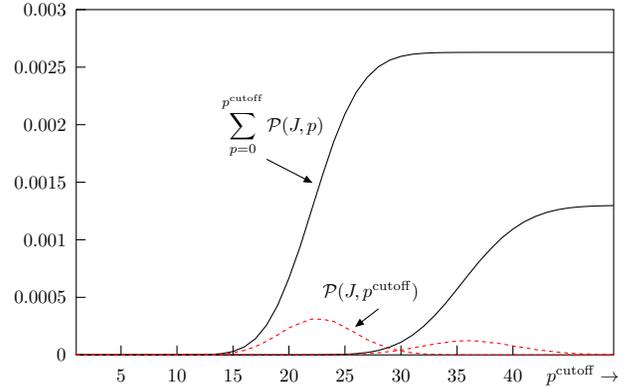}
\caption{Plot of the summed probability, in $U(2)$, to find a state
  with zero or more $\hat{O}_2$ operators in $|c\rangle$, as a
  function of the maximum number of operator insertions~$p^\text{cutoff}$
  in~\eqn{expectations}. Equivalently, this is the total probability
  to find a singlet state in the coherent state. The two continuous
  lines correspond to two different values of the coupling
  constant. The dashed lines are the individual terms that make up the
  sum. Observe that these curves never reach one, which shows that
  there are still many states in the coherent state which are
  non-singlets. }
\label{f:sumind}
\end{center}
\end{figure}

With the correct norms of the multi-particle states at hand, one now
obtains sensible results for the sums~\eqn{e:numbers}
and~\eqn{e:energies}. An example in U(2) (which is rather trivial
because there is in this case only one independent single-trace
operator) is plotted in figure~\ref{f:sumind}. Note once more that the
numbers plotted here are much smaller than one. This is because the
norm ${\cal C}$ which multiplies all of these results is the norm of
the non-singlet coherent state, and the number of singlets in it is
much smaller than the total number of states. Note, however, that
since the norm of the coherent state appears in all probabilities as
an overall (identical) factor, its absolute value is irrelevant when
considering the ratios of emitted energies or ratios of numbers of
particles.  See appendix~\ref{s:singletprojections} for a more
explicit explanation.

Having resolved the computation of the exact norms of states, we can
now finally compute the energy distribution in the outgoing state of a
more interesting example. For practical reasons, we will restrict
ourselves to the~U(4) case, for which there are only two operators
which create physical states (using only the creation operator for the
lowest-lying spherical harmonics). These
operators are~$\Tr\big((a^\dagger)^2\big)$ and
$\Tr\big((a^\dagger)^4\big)$.\footnote{The restriction to the
zero-mode of the scalar field is motivated by the full sphaleron
solution of the earlier sections, which only turns on the lowest
spherical vector harmonics.  Naturally, in the full $U(4)$ there are
also operators of the form $\Tr(D_\mu\phi D_\nu\phi)$. However, in the
oscillator picture these are turned on by the oscillators that create
the higher spherical tensorial harmonics.}  The proper linear
combinations of these operators are
\begin{equation}
\begin{aligned}
\hat{O}_2^\dagger &= \Tr( a^\dagger a^\dagger )\,,\\[1ex]
\hat{O}_4^\dagger &= \Tr( a^\dagger a^\dagger a^\dagger a^\dagger ) -
\frac{2N^2 + 1}{N(N^2+2)} \Tr( a^\dagger a^\dagger ) \Tr( a^\dagger
a^\dagger )\,.
\end{aligned}
\end{equation}
% V2: refined the statement about orthogonality of the states.
These lead to $\langle \hat{O}_4\,|\, \hat{O}_2
\hat{O}_2\rangle=0$. Multi-particle states will
generically not be orthogonal (see also footnote~\ref{f:ortho}), but
in our case this turns out to be far less important than the
$1/N^2$~corrections to the norms. We will for simplicity use a
classical configuration for which
\begin{equation}
\frac{\eta_4}{N} = \left(\frac{\eta_2}{N}\right)^2 = \frac{\eta}{N}\,,
\end{equation}
where the $\eta_J$ are defined in~\eqn{argu}. Closer inspection of the
coherent state of the sphaleron given in~\eqn{coherent} shows that the
expectation values of e.g.~the $\Tr(F_{mn} F^{mn})$ and~$\Tr(F_{m
n}F^{m n} F_{rs} F^{rs})$ states are similarly related.

The energy radiated into $O_2$ and $O_4$ particles can be computed
using formula~\eqn{e:energies}, summed over a suitably large range of
values for $p_2$ and $p_4$. In our particular case, this formula
reduces to
%\begin{widetext}
\begin{equation}
\label{e:cutoffsum}
\begin{aligned}
E&(J,p^{\text{\maxp}}_2, p^{\text{\maxp}}_4) =\\[1ex]
  &\sum_{p_2=0}^{p_2^{\text{\maxp}}} \sum_{p_4=0}^{p_4^\text{\maxp}}
  \left|\frac{\eta_2^2}{\lambda^2}\right|^{p_2}
  \left|\frac{\eta_4^2}{\lambda^4}\right|^{p_4}
  \frac{J p_J}{2^{p_2} 4^{p_4}} \\[1ex]
&\times  \frac{{\cal C}^2}{\langle 0|\, (\hat{O}_2)^{p_2} (\hat{O}_4)^{p_4}\,
                  (\hat{O}_4^\dagger)^{p_4} (\hat{O}_2^\dagger)^{p_2}\,
                  |0\rangle\,\langle c|c\rangle}\,.
\end{aligned}
\end{equation}
%\end{widetext}
and the maximum values of $p_2$ and $p_4$ which are included in the
sum should be taken sufficiently large as to include at least the
maximum term in the sum. This requirement is indeed met in our
numerical approach.
\psfrag{10}{$\scriptstyle 10$}
\psfrag{20}{$\scriptstyle 20$}
\psfrag{40}{$\scriptstyle 40$}
\psfrag{60}{$\scriptstyle 60$}
\psfrag{-20}{$\scriptstyle ~-20$}
\psfrag{-40}{$\scriptstyle ~-40$}
\psfrag{-60}{$\scriptstyle ~-60$}
\psfrag{-80}{$\scriptstyle ~-80$}
\psfrag{0}{$\scriptstyle 0$}
\psfrag{p2}{$\scriptstyle p_2^\text{\maxp}$}
\psfrag{p4}{$\scriptstyle p_4^\text{\maxp}$}
\begin{figure*}
\begin{center}
\includegraphics*[width=.6\textwidth]{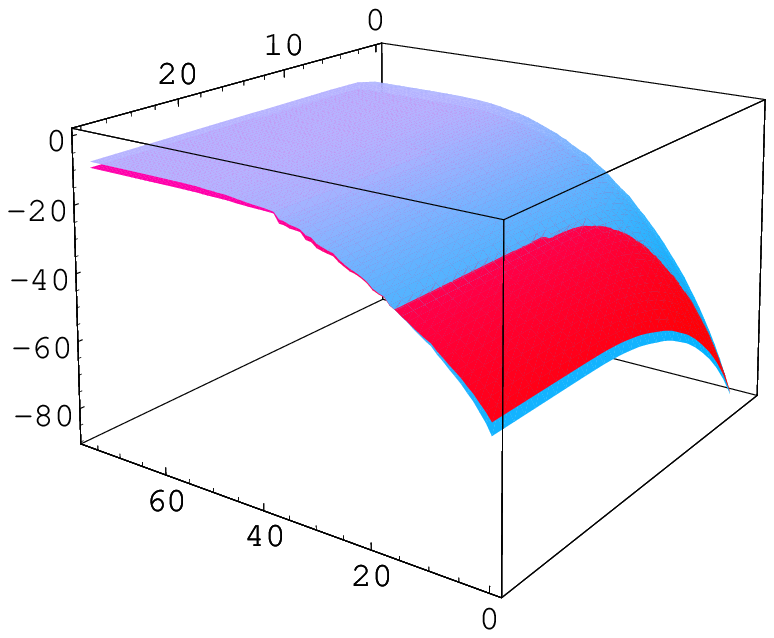}
\makebox[0pt]{\hspace{-1.3cm}\raisebox{2.63cm}{\includegraphics*[height=5.2cm]{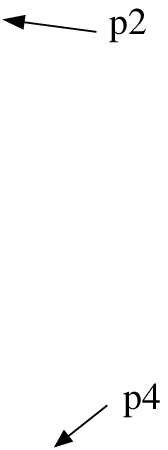}}}
\caption{Successive approximations to the logarithm of the total
energy radiated in the $J=2$ particles (light, blue surface) and $J=4$
particles (dark, red surface). The $x$ and $y$ axes label the maximum
value of $p_2$ and $p_4$ in the sum~\eqn{e:cutoffsum}. The values
asymptote to the full result in the upper left corner of the
graph. The curves in the $x-z$ and $y-z$ plane are similar to the
continuous lines in figure~\ref{f:sumind}. While the present plot
shows energies, qualitatively similar plots are obtained for the
particle numbers.}
\label{f:energies}
\end{center}
\end{figure*}%
We have computed the ratio of energies in the $J=2$ and $J=4$
particles using successive approximations of~\eqn{e:cutoffsum}, for
larger and larger $p^{\text{\maxp}}_2$ and $p_4^{\text{\maxp}}$,
\begin{equation}
\label{e:aim}
\lim_{\substack{\\p^{\text{\maxp}}_2\rightarrow\infty\\[.5ex]p^{\text{\maxp}}_4\rightarrow\infty}}
\frac{E(4,p^{\text{\maxp}}_2,p^{\text{\maxp}}_4)}{E(2,p^{\text{\maxp}}_2,p^{\text{\maxp}}_4)} =: R(\eta^2/\lambda^2)
\end{equation}
for a range of couplings. A typical example is plotted in
figure~\ref{f:energies}. One clearly sees that the asymptotic value of
the ratio~\eqn{e:aim}, given by the exponent of the asymptotic height
difference between the two surfaces, is smaller than one. We therefore
conclude that our calculation predicts that higher-energy states in
the decay product are suppressed with respect to the lower-energy
ones. This is in qualitative agreement with alternative calculations
of this decay process~\cite{Lambert:2003zr}.

It would be very interesting to extend our analysis to higher-rank
gauge groups, perhaps by obtaining an analytic expression for the
norms of the states. For~$N>4$, there are more than two gauge singlet
states, and it becomes possible to determine the suppression factor as
a function of the energy in more detail. We leave this for future
investigations.

\section{Discussion and open issues}

We have presented the formalism to analyse the decay of unstable
D-branes in the~$\text{AdS}_5 \times S^5$ background by considering
the dual gauge theory. Our results show qualitative agreement with
previous work on D-particle decay, and our paper provides the basis
for further study of non-perturbative dynamical features of the
correspondence. Let us conclude by describing a number of open issues
and possible extensions of our work.

% V3: references added (below) for group character methods.
One obvious way to improve on our results would be to determine
analytical expressions for the norms required in
section~\ref{s:U4results} (using the construction of states in terms
of group
characters~\cite{Balasubramanian:2001nh,Corley:2001zk,Kristjansen:2002bb}). This
would allow one to extend the results obtained there to large values
of~$N$. We expect that already for the~U(6) model it should be
possible to get evidence that the observed suppression of the decay
products with their mass is actually exponential. It would be
interesting to see whether this suppression is strong enough to
compete with the Hagedorn growth of the multiplicity of states at high
mass levels. Obtaining these results should enable one to check
whether the total energy emitted in the decay product is finite or
not.

The flat space string calculation of~\cite{Lambert:2003zr} and the
matrix model calculation of~\cite{McGreevy:2003kb} both obtained an
infinite total energy for the final decay product. One might think
that the reason for such behavior is that there is non-trivial
back-reaction of the radiated closed strings on the boundary state,
which has not been taken into account. However, it was argued
in~\cite{Klebanov:2003km,Sen:2003iv,Sen:2004zm} that the time
dependent boundary state of~\cite{Sen:2002nu} already contains the
full information about the closed string sector into which it is going
to decay.  Instead, the reason for the divergences observed
in~\cite{Lambert:2003zr,McGreevy:2003kb} has been attributed to the
fact that the coherent state of the unstable brane has an infinite
spread in energy (in the fermionic description it corresponds to a
sharply localised fermion in position space).

We believe that, whatever the reason for the observed divergence, the
emitted energy calculated from our coherent state should be finite.
In our setup there is no issue of backreaction, since there is no
separation between the source and the ``remainder'' in our system. For
the construction of the coherent state we have used a solution of the
full, non-linear Yang-Mills equations of motion.  Also, as one can
check, the coherent state thus constructed has finite spread both in
momentum and position space, hence avoiding the aforementioned
problem.

To make a link of our work to the comments
of~\cite{Sen:2003iv,Sen:2004zm} and as a comparison to the matrix
model, let us note that the classical tachyon evolution is governed by
the reduced Yang-Mills action~\eqn{e:reducedYM} (or, to be more
precise, to a similar reduction based on more general gauge group
embeddings of the type~\eqn{e:sunsu2}). One might hope that this
action is dual to the open string field theory on decaying D-particles
in the bulk of~AdS.  However, this requires further analysis.

As we have explained, due to the non-perturbative nature of the
initial sphaleron configuration, the computation of the decay product
requires information from a regime in which
both~\mbox{$N\rightarrow\infty$} as well as the number of particles
~$p\rightarrow\infty$. Understanding this double limit might
circumvent the need to calculate the state norms exactly when
calculating the energy distribution in the final state. Finally, it
would be interesting to understand how quantum corrections can be
incorporated into our formalism, in order to see how much they
influence the qualitative characteristics of the decay product.

\section*{\small Acknowledgements}

We would like to thank Gleb Arutyunov, Justin David, Charlotte
Kristjansen and Jan Plefka for discussions and especially Rajesh
Gopakumar, Stefano Kovacs, Shiraz Minwalla and Ashoke Sen for
comments on a draft of this paper.

\vfill\eject
\section{Appendix}
\subsection{$S^2$ operator--state correspondence}
\label{a:s2opstate}

We will here consider the operator--state correspondence in the
context of a simple~$S^2 \times \mathbb{R}$ example. Using the
procedure outlined in the main text, let us construct all states
corresponding to set of operators
\begin{equation}
\label{e:operators}
\begin{aligned}
\hat O_1 &:= \Tr\Big( \partial_{(\mu} \chi \partial_{\nu)} \chi \Big) -
\frac{1}{d} g_{\mu\nu} 
\Tr\Big( \partial_\rho \chi \partial^\rho \chi\Big)
\,,\\[1ex]
\hat{O}_2 &:= \Tr\Big( \partial_\rho \chi \partial^\rho
\chi\Big)\,,\\[1ex]
\hat{O}_3 &:= \Tr\Big( \chi\partial_\mu\partial_\nu\chi\Big) \,.
\end{aligned}
\end{equation}
Using the counting of states introduced by
Sundborg~\cite{Sundborg:1999ue} and Polyakov~\cite{Polyakov:2001af}
(see also Aharony et al.~\cite{Aharony:2003sx}) we see that the total
number of states created by these operators is
\begin{equation}
\label{e:shiraz}
\begin{aligned}
\partial_\mu\chi \partial_\nu\chi &:\quad \frac{3\cdot4}{2} = 6 \quad\text{states}\,,\\
\chi\partial_\mu\partial_\nu\chi  &:\quad \frac{3\cdot4}{2}-1 = 5\quad\text{states}\,,
\end{aligned}
\end{equation}
giving in total $11$ states. This means that the operators
(\ref{e:operators}) also create $11$ states, since they are all
possible operators one can build out of building blocks
(\ref{e:shiraz}). Let us first count the number of states created by the
operator $\hat{O}_2$. The operator has dimension~$\Delta=3$, hence we can
decompose it onto the various states as follows
\begin{widetext}
\begin{equation}
\label{e:O2decompose}
\lim_{\tau\rightarrow-\infty}\int_{S^2} e^{-3\tau} \sum_{l'',m''}
Y_{l''m''} \,
\partial_\rho \Big(\sum_{l,m} e^{\frac{1}{2}(2l+1)\tau} Y^*_{lm} a^\dagger_{lm}\Big)
\partial^\rho \Big(\sum_{l',m'} e^{\frac{1}{2}(2l'+1)\tau} Y^*_{l'm'}
a^\dagger_{l'm'}\Big)|0\rangle\, . 
\end{equation}
Note that while the spherical harmonics which figure in the expansion
of the field $\chi$ are on-shell, the spherical harmonics $Y_{l''m''}$
onto which we project need not be on-shell.  Clearly the options for
$(l,l')$ in (\ref{e:O2decompose}) are $(2,0)$, (i.e. $(0,2)$) and
$(1,1)$.  The explicit expressions before integration are (${\cal
  N}_{i,j}$ are normalisation constants)
% V2: explained normalisation constants,
%     added vacuum states on the rhs.
\begin{equation}
\label{e:st11}
\begin{aligned}
\partial_\mu \chi \partial^\mu \chi \Big|_{(l,l')=(0,2)} |0\rangle &=
 \frac{5}{8 \sqrt{\pi}}\sum_{m'} Y^*_{2m'} \,
      a^\dagger_{00} a^\dagger_{2m'} |0\rangle \,, \\
\partial_\mu\chi\partial^\mu\chi\Big|_{(l,l')=(1,1)} |0\rangle
 &= \frac{5}{4\sqrt{\pi}} 
   \Bigg[ \begin{aligned}[t]
         &(a^\dagger_{1,-1})^2 {\cal N}_{1,-1}^2
              \Big(   \frac{5}{4} \sin^2\theta \Big) e^{2i\phi}
        + (a^\dagger_{1,1})^2 {\cal N}_{1,1}^2 
              \Big(   \frac{5}{4} \sin^2\theta \Big) e^{-2i\phi} \\
        +&(a^\dagger_{1,0})^2 {\cal N}_{1,0}^2
              \Big(   \frac{9}{4} \cos^2\theta \Big) \\
        +&\,a^\dagger_{1,1}  a^\dagger_{1,-1} {\cal N}_{1,1} {\cal N}_{1,-1}
              \Big( - 4 - \frac{5}{2} \sin^2\theta  \Big) \\
        +&\,a^\dagger_{1,-1}  a^\dagger_{1,0} {\cal N}_{1,-1} {\cal N}_{1,0}
              \Big(  \frac{5}{2} \sin\theta \cos\theta \Big) e^{i\phi} \\
        +&\,a^\dagger_{1,1}  a^\dagger_{1,0} {\cal N}_{1,1} {\cal N}_{1,0}
              \Big( -4 \sin\theta\cos\theta \Big) e^{-i\phi} \Bigg] |0\rangle\,.
        \end{aligned}
\end{aligned}
\end{equation}
\end{widetext}
% V2: cleaned up the text below
We see that projection onto the $Y_{l''m''}$ modes with $l''=2$ would
give five non-zero projections (the $a^\dagger_{1,1} a^\dagger_{1,-1}$
and $(a^\dagger_0)^2$ come together), which is too many.  The correct
procedure is to use the lowest harmonic $l''=0$, in which case only
one state (the second and third lines in (\ref{e:st11})) is selected.
If one repeats similar exercises with the operators $\hat{O}_1$ and
$\hat{O}_3$, one obtains states that are not orthogonal to states
obtained from $\hat{O}_2$ unless one uses the appropriate lowest lying
tensor spherical harmonic. To see this we need to use the tensor
harmonics on $S^2$.
%constructed in \cite{}.
There are four types of 
lowest-lying 2-tensor spherical
harmonics:
\begin{alignat}{3}
\eta_{ab}^{(lm)} &= Y^{(lm)} g_{ab}\,,  & \quad l\geq 0\,,\\[1ex]
\psi_{ab}^{(lm)} &= Y^{(lm)}_{;ab} + \tfrac{1}{2}l(l+1)\,, & \quad l\geq 2\,,\\[1ex]
\chi_{ab}^{(lm)} &= Y^{(lm)} \epsilon_{ab}\,, & \quad l\geq 0\,,\\[1ex]
\phi_{ab}^{(lm)} &= \tfrac{1}{2}\big( \phi_{a;b}^{(lm)} +
\phi_{b;a}^{(lm)}\big)\,, & \quad l\geq 2\, ,
\end{alignat}
where $\phi_{a}^{(lm)}$ is a vector spherical harmonic, given by
$\phi_{a}^{(lm)} = \epsilon_{a}{}^b Y^{(lm)}_{,b}$.
The $\eta$ mode is a pure trace so it does not contribute when
contracted with $\hat{O}_1$ or $\hat{O}_3$ and the equations of motion
for $\chi$ are used. Furthermore, the $\chi$ mode is anti-symmetric,
so it also leads to a zero. Thus we find, by explicitly contracting
$\psi$ and $\phi$ with the two operators, multiplying with
$e^{-3\tau}$ and taking the $\tau\rightarrow-\infty$ limit, and
finally integrating over~$S^2$, that
\begin{equation}
\begin{aligned}
\int_{S^2} \psi_{ab}^{(l=2,m=-2)} \hat{O}_1^{ab}\, {\rm d}\Omega 
  &\sim \big( a^\dagger_{1,1}\big)^2 |0\rangle\,,\\[1ex]
\int_{S^2} \phi_{ab}^{(l=2,m=-2)} \hat{O}_1^{ab}\, {\rm d}\Omega &= 0\,,\\[1ex]
\int_{S^2} \psi_{ab}^{(l=2,m=-2)} \hat{O}_3^{ab}\, {\rm d}\Omega 
  &\sim a^\dagger_{2,2} a^\dagger_{0,0} |0\rangle\,,\\[1ex]
\int_{S^2} \psi_{ab}^{(l=2,m=-2)} \hat{O}_3^{ab}\, {\rm d}\Omega &= 0\, ,
\end{aligned}
\end{equation}
and with similar expressions for the other~$4$ labels~$(2,1)$,
$(2,0)$, $(2,-1)$ and $(2,-2)$. Note however, that all these
expressions involve \emph{different bilinears of operators
$a^\dagger$} and hence are automatically orthogonal.

In summary, we thus find that the operators $\hat{O}_1$ and
% V2: typo corrected (O_2 -> O_3)
$\hat{O}_3$ create $5$ states each, while operator $\hat{O}_2$
corresponds to a single state, altogether giving a total of~$11$
states as required by~\eqn{e:shiraz}.

\subsection{Singlet projections}
\label{s:singletprojections}

We will here use a simple example to show how the smallness of the
expectation values plotted in e.g.~figure~\ref{f:sumind} can be
understood.  As mentioned in the main text, the crucial reason is that
the coherent state used to make these plots was the state~$|c\rangle$
(rather than the state~\eqn{e:project}), and this state
contains both singlet and non-singlet states.  We have argued that the
suppression in figure~\ref{f:sumind} is determined by the number of
singlets versus the number of non-singlet states in~$|c\rangle$. To
illustrate this, we will here use the~$J=2$ operator in U(2), for a
non-abelian scalar. There are four elementary operators
$\hat{a}^\dagger,\ldots \hat{d}^\dagger$,
\begin{equation}
\hat{A}^\dagger = \begin{pmatrix} \hat{a}^\dagger & \hat{b}^\dagger \\ 
                          \hat{c}^\dagger & \hat{d}^\dagger \end{pmatrix}\,.
\end{equation}
and each of these satisfies a canonical commutation relation with its
conjugate. For the classical field, let us take the simple example of
\begin{equation}
A_{\text{class.}} = \begin{pmatrix} \eta & 0 \\ 0 & 0 \end{pmatrix}\,.
\end{equation}
This implies that the coherent state is given by
\begin{equation}
|c\rangle = {\cal C}\, \exp\Big( \frac{1}{g^2} \Tr\big(
 A_{\text{class.}} \hat{A}^\dagger \big) \Big) |0\rangle 
 = {\cal C}\, \exp\Big(\frac{1}{g^2} \eta \hat{a}^\dagger \Big) |0\rangle\,.
\end{equation}
The correct normalisation constant is thus
\begin{equation}
{\cal C}^2 = \exp\big(-\frac{\eta^2}{g^2}\big)\,.
\end{equation}
We always compute projections of the coherent state onto
gauge-singlet states. Let us consider the~$p=1$ case,
\begin{equation}
\frac{\Big| \big\langle 0 \big| \Tr(\hat{A}^2) \big| c\big\rangle \Big|^2}{%
  {\big<0\big| \Tr(\hat{A}^2)\, \Tr((\hat{A}^\dagger)^2) \big|0\big\rangle\,
   \big\langle c \big| c\big\rangle}}\,.
\end{equation}
The norm in the denominator equals $2 N^2 g^4 = 8 g^4$. This gives, if
one adds the trivial~$p=0$ term,
\begin{equation}
{\cal P}(J=2) = {\cal C}^2 \Big( 1 + \frac{\eta^4}{8\, g^4} +
\ldots\Big)
 = \frac{\Big( 1 + \frac{\eta^4}{8\, g^4} + \ldots\Big)}{%
         \Big( 1 + \frac{\eta^2}{g^2} + \frac{\eta^4}{2\,g^4} + \ldots\Big)}\,.
\end{equation}
Here we have expanded ${\cal C}^2$ in the second step. The fact
that~${\cal P}(J=2)$ comes out smaller than~$1$ has two
reasons. Firstly, the odd powers of $\eta^2/g^2$ are manifestly absent
from the numerator, since they correspond to the states with odd
powers of the operator~$\hat{a}$ and are hence manifestly
non-singlets.  Secondly, the coefficient of~$\eta^4/g^4$ in the
numerator is only~$1/8$, as compared to the~$1/2$ in the
denominator. This is because out of all quadratic operators that
involve the operator $\hat{a}$ in some combination with the operators
$\hat{a}\ldots \hat{d}$, only one is a trace operator. We are focusing
on the operators that necessarily involve~$\hat{a}$, since all other
operators are absent in the coherent state.  In the sector which
contains $\hat{a}^2$, $\hat{b}\hat{c}$ and $\hat{d}^2$, there are three of those,
\begin{equation}
\hat{O}_1 = \hat{a}^2 + 2 \hat{b}\hat{c} + \hat{d}^2 \,,\quad
\hat{O}_2 = \hat{a}^2 - 2 \hat{b}\hat{c} + \hat{d}^2 \,,\quad
\hat{O}_3 = \hat{a}^2 - \hat{d}^2\,.
\end{equation}
As given here, these are orthogonal. However, only $\hat{O}_1$ is a
single-trace operator.

If we also compute the projection of the coherent state onto
$\hat{O}_2$ and $\hat{O}_3$, and add these probabilities to the one
found for $\hat{O}_1$, we get
\begin{equation}
{\cal P}(J=2) = {\cal C}^2 \Big( 1 + \frac{\eta^4}{g^4}\Big[
  \frac{1}{8} + \frac{1}{8} + \frac{1}{4} \Big] + \ldots \Big)\,.
\end{equation}
Now the $1/2$ precisely matches the $1/2$ in ${\cal C}^2$.

Note that this issue becomes more and more serious as we go up in the
number of operator insertions. The very small numbers as presented in
section~\ref{s:U4results} thus arise because these only count
multi-particle singlet states, as opposed to generic multi-particle
states.

\subsection{Geometrical expressions}

In this section we collect some intermediate results of the
calculations and some useful formulae that were used in the main text.
The coordinates which we use on $\mathbb{R}\times S^3$ are related to
Cartesian coordinates via
\begin{equation}
\begin{aligned}
z_1 &= x^0 + i x^1  \\[1ex] r \cos\left(\frac{\theta}{2}\right)
    &=\left(\cos\left(\frac{\phi + \psi}{2}\right)+ i \sin\left(\frac{\phi + \psi}{2}\right)\right) \,, \\
z_2 &= x^3 + i x^4  \\[1ex] r \cos\left(\frac{\theta}{2}\right)
    &=\left(\cos\left(\frac{\phi - \psi}{2}\right)+ i \sin\left(\frac{\phi + \psi}{2}\right)\right)\,,
\end{aligned}
\end{equation}
after which we perform a conformal rescaling to obtain the metric
\begin{equation}
\label{e:RS3metric}
{\rm d}s^2 = -{\rm d}t^2 + \frac{R^2}{4}\Big({\rm d}\theta^2 + {\rm d}\psi^2 
                     + {\rm d}\phi^2 + 2\, \cos \theta \, {\rm d} \psi
                     {\rm d} \phi \Big)\, .
\end{equation}
The coordinate ranges are given by $\theta\in[0,\pi\rangle$,
$\phi\in[0,2\pi\rangle$ and $\psi\in[0,4\pi\rangle$. The volume of the
$S^3$ part is thus computed to be $\vol(S^3)=2\pi^2 R^3$. The inverse
metric is
\begin{equation}
g^{\theta\theta} = \frac{4}{R^2} \, \quad g^{\phi\phi} = g^{\psi\psi} = \frac{4}{R^2\,\sin^2(\theta)} \, \quad g^{\psi \phi} = - \frac{4 \cos \theta}{R^2\,\sin^2 \theta}
\end{equation}
and the connection
\begin{equation}
\begin{aligned}
\Gamma^{\theta}_{\phi \psi} &= \frac{1}{2}\sin \theta \,,\\[1ex]
\Gamma^{\phi}_{\phi \theta} &= \Gamma^{\psi}_{\psi \theta} = \frac{\cos \theta}{2 \sin \theta}  \,,\\[1ex]
\Gamma^{\phi}_{\psi \theta} &= \Gamma^{\psi}_{\phi \theta} = - \frac{1}{2 \sin \theta} 
\end{aligned}
\end{equation}
The gauge potential (\ref{ansatz}) in coordinates $(t, \theta, \phi, \psi)$ reads 
\begin{equation}
\begin{aligned}
A_t &= 0 \,,\\[1ex]
A_{\theta} &= \frac{i}{2} f(t) (\cos\phi \,\sigma_2 - \sin \phi \, \sigma_3) \,, \\
A_{\phi} &= \frac{i}{2} f(t) \sigma_1 \,,\\[1ex]
A_{\psi} &= \frac{i}{2} f(t)(\cos\theta \, \sigma_1 + \sin\theta
\sin\phi \, \sigma_2 + \sin \theta \cos \phi \, \sigma_3 ) \, .
\end{aligned}
\end{equation}
The field strength is defined by 
\begin{equation}
\label{fs}
F_{\mu\nu} = \partial_{\mu} A_{\nu} - \partial_{\nu} A_{\mu} - [A_{\mu}, A_{\nu}] \, , 
\end{equation}
and the corresponding gauge transformations are 
\begin{equation} 
\label{gtr}
A_{\mu} \rightarrow \Lambda A_{\mu} \Lambda^{\dagger} - \Lambda \partial_{\mu} \Lambda^{\dagger} \, .
\end{equation}
For the sphaleron configuration (\ref{ansatz}), the field strenghts are given by 
\begin{equation}
\begin{aligned}
F_{\theta \phi} &= B (\sin \phi \sigma_2 + \cos \phi \sigma_3) \,, \\
F_{\theta \psi} &= B (- \sin \theta \sigma_1 + \sin \phi \cos \theta \sigma_2 + \cos \theta \cos \phi \sigma_3 ) \,, \\
F_{\phi \psi} &= B (\sin \theta \cos \phi \sigma_2 - \sin \theta \sin \phi \sigma_3) \, \\
F_{0\theta} &= \frac{\dot{f}}{f} A_\theta \, \quad F_{0\psi} = \frac{\dot{f}}{f}A_\psi \, 
\quad F_{0\phi} = \frac{\dot{f}}{f} A_\phi\,, \\
B &\equiv \frac{i f}{2} (1- f)\,.
\end{aligned}
\end{equation}
% and with upper indices
% \begin{eqnarray}
% F^{\theta\phi} &=& 16 B (\cot \theta \sigma_1 + \sin \phi \sigma_2 + \cos \phi \sigma_3) \, \nonumber \\
% F^{\theta\psi} &=& - \frac{16B}{\sin \theta} \sigma_1 \, \nonumber \\
% F^{\phi\psi} &=& \frac{16B}{\sin^2 \theta} (\sin \theta \cos \phi \sigma_2 - \sin \theta  \sin \phi \sigma_3)
% \end{eqnarray}
The lowest-order vector spherical harmonics are related to the
canonically normalised left-invariant one-forms as
\begin{widetext}
\begin{equation}
\begin{aligned}
Y_{\tfrac{1}{2}(0,0)\tfrac{1}{2}}  &= 
   \frac{i}{2}\frac{1}{\sqrt{\vol(S^3)}} \Sigma^3\,,
&
\Sigma^1 &= \sqrt{2\vol(S^3)} i \Big( Y_{\tfrac{1}{2}(1,0)\tfrac{1}{2}} + 
                                      Y_{\tfrac{1}{2}(-1,0)\tfrac{1}{2}} \Big)\,,\\[1ex]
Y_{\tfrac{1}{2}(-1,0)\tfrac{1}{2}} &=
   \frac{1}{2\sqrt{2}}\frac{1}{\sqrt{\vol(S^3)}}\Big( -\Sigma^2 - i \Sigma^1 \Big)\,,\quad&
\Sigma^1 &= \sqrt{2\vol(S^3)} i \Big( Y_{\tfrac{1}{2}(1,0)\tfrac{1}{2}} + 
                                      Y_{\tfrac{1}{2}(-1,0)\tfrac{1}{2}} \Big)\,,\\[1ex]
Y_{\tfrac{1}{2}(1,0)\tfrac{1}{2}}  &=
   \frac{1}{2\sqrt{2}}\frac{1}{\sqrt{\vol(S^3)}}\Big( +\Sigma^2 - i
   \Sigma^1 \Big)\,,\quad&
\Sigma^3 &= -2i\sqrt{\vol(S^3)} Y_{\tfrac{1}{2}(0,0)\tfrac{1}{2}} \,.
\end{aligned}
\end{equation}
Here the left-invariant one-forms are given by
\begin{equation}
\begin{aligned}
\Sigma^1 &= \hphantom{-}\cos\psi\, {\rm d}\theta + \sin\psi\sin\theta\, {\rm
d}\phi\, ,\\
\Sigma^2 &= -\sin\psi\, {\rm d}\theta + \cos\psi\sin\theta\, {\rm
d}\phi \, ,\\
\Sigma^3 &= \hphantom{-}\cos\theta\, {\rm d}\phi + {\rm d}\psi\, ,
\end{aligned}
\end{equation}
\end{widetext}
For the explanation of indices, see formula (\ref{indices}).

\begin{small}
\bibliography{sphaleron_revtex}
%\bibliography{kasbib}
\end{small}
\end{document}